\documentclass[aps,pra,twocolumn,showpacs,floatfix,groupedaddress,superscriptaddress]{revtex4-1}
\usepackage{graphicx,graphics,times,amssymb,amsmath,amsfonts,bm,bbm}

\begin{document}

\title{Effect of weak magnetic field on quantum spin correlation in an $s$-wave superconductor}

\author{R. Afzali}
\affiliation{Department of Physics, K. N. Toosi University of Technology, Tehran 15418, Iran}
\email{afzali@kntu.ac.ir}

\author{A. T. Rezakhani}
\affiliation{Department of Physics, Sharif University of Technology, Tehran, Iran}

\author{H. R. Alborznia}
\affiliation{Department of Physics, Azad University--Tehran-Central, Tehran, Iran}

\begin{abstract}
We study the effect of a weak magnetic field on the state of the spins of a pair of electrons forming a Cooper pair in an $s$-wave superconductor. By perturbatively solving time-dependent Bogoliubov equations up to first order, we obtain the two-particle Green function, and whence the two-electron spin-space density matrix. It appears that up to first order approximation, the spin state of a Cooper pair retains the form of a Werner state. This state is then examined in the sense of quantum correlations, i.e., quantum discord. 
\end{abstract}


\maketitle


In quantum manybody or strongly correlated systems, quantum correlations are responsible for a host of interesting physical phenomena. Specifically, quantum entanglement is known to be at the heart of quantum phase transitions and a major player in quantum information science and its applications \cite{ref1,ref2,Nielsen:book}. It is thus of fundamental and practical importance to analyze quantum correlations generated in quantum manybody systems. As an example, recently it was shown \cite{ref3} that in a non-interacting electron gas and at low temperatures, spin-spin entanglement behaves quadratically with temperature. Additionally, two electron spins forming Cooper pairs in an $s$-wave superconductor has been shown to be a Werner state, whose entanglement may survive within a coherence length $\xi_0$ \cite{ref4}. In quantum correlated systems such as extended Hubbard model, whose states are amenable to exact or approximate calculations, the role of entanglement (i.e., quantum correlations) can be identified more explicitly \cite{Hubbard}.

In this Brief Report, we consider an $s$-wave superconductor subjected to a weak magnetic field. Of our special interest is the state of a Cooper pair in such a system and how applying a field can alter quantum correlations between the spins of the electrons in this pair. To this end, we first obtain the density matrix of the system by employing the Green functions and solving the related Bogoliubov equations perturbatively. It is shown that this results in a Werner form for the spin-space density matrix. Additionally, we calculate quantum correlations---exactly, quantum discord \cite{Zurek}---of this state, from whence one can see how applying a weak magnetic field modifies quantum correlations in the level of a Cooper pair. 

%

The Hamiltonian of a generic superconductor subjected to weak magnetic field is as follows \cite{ref14,ref15,ref16}:
\begin{eqnarray}  
H_{\text{eff}} &= &\sum _{\sigma }\int \Psi _{\sigma }^{\dag} (\mathbf{r}) \bigl[H_{\text{e}}(\mathbf{r}) + U(\mathbf{r})\bigr]  \Psi _{\sigma } (\mathbf{r})~\mathrm{d}^3\mathbf{r}\\
&&~+ \int \bigl[\Delta(\mathbf{r}) \Psi _{\uparrow }^{\dag} (\mathbf{r})\Psi _{\downarrow }^{\dag} (\mathbf{r})+  \Delta^*(\mathbf{r})\Psi _{\downarrow } (\mathbf{r})\Psi _{\uparrow } (\mathbf{r})\bigr]~\mathrm{d}^3\mathbf{r},\nonumber
\label{eq-1_}
\end{eqnarray} 
where $\Psi^{\dag}$ ($\Psi$) is the fermionic field creation (annihilation) operator, $H_{\text{e}}(\mathbf{r})\equiv \frac{1}{2m}[i\hbar \nabla -e\textbf{A}(\textbf{r})/c]^{2} -\mu$, $\Delta(\mathbf{r})=V(\mathbf{r})\langle \Psi _{\uparrow } (\mathbf{r})\Psi _{\downarrow }(\mathbf{r})\rangle $ is the energy gap [which is assumed to be constant (i.e., $\mathbf{r}$-independent) in $s$-wave superconductors], $U(\mathbf{r})= -V(\mathbf{r})\langle \Psi_{\uparrow}^{\dag}(\mathbf{r}) \Psi_{\uparrow}(\mathbf{r}) \rangle$ is the mean (one-electron) potential, $V(\mathbf{r})$ is a potential characterizing electron-electron interactions, $\mathbf{A}(\mathbf{r})$ is a vector potential, $\mu$ is the chemical potential, and $\sigma\in\{\uparrow,\downarrow\}$. Here, $\left\langle \cdots\right\rangle$ implies a combined quantum mechanical and statistical average \cite{ref16}. Note, however, that later (to be specified) in this Report, we shall work only at zero temperature, wherein the average reduces to $\langle \psi_{\text{BCS}}|\ldots|\psi_{\text{BCS}}\rangle$ ($|\psi_{\text{BCS}}\rangle$ is the Bardeen-Cooper-Schrieffer state).

This Hamiltonian can be exactly diagonalized by using the following Bogoliubov transformation: 
\begin{eqnarray}
&\Psi _{\uparrow } (\mathbf{r})=\sum _{n} u_{n} (\mathbf{r})\gamma _{n\uparrow } - v^*_{n}(\mathbf{r}) \gamma _{n\downarrow }^{\dag} ,\\   
&\Psi _{\downarrow } (\mathbf{r})=\sum _{n} u_{n}(\mathbf{r}) \gamma _{n\downarrow } + v^*_{n} (\mathbf{r}) \gamma _{n\uparrow }^{\dag},                                       
\label{eq-2_}
\end{eqnarray}
as $H=\sum _{n\sigma}\varepsilon_{n} \gamma_{n\sigma }^{\dag} \gamma _{n\sigma}$, in which $\gamma _{n\sigma}^{\dag}$ ($\gamma _{n\sigma}$) are some bosonic creation (annihilation) operators satisfying $\{\gamma_{n\sigma},\gamma_{n'\sigma'}\}=\{\gamma _{n \sigma}^{\dag} ,\gamma_{n' \sigma'}^{\dag}\}=0$ and $\{\gamma _{n\sigma} (\mathbf{r}),\gamma_{n' \sigma'}^{\dag} (\mathbf{r}')\}=\delta _{\sigma\sigma'}\delta_{nn'} \delta (\mathbf{r}-\mathbf{r}')$. Here, $\varepsilon_{n}$ is the corresponding energy spectrum, $u_{n} (\mathbf{r})$ and $v_{n}(\mathbf{r})$ are the Bogoliubov coefficients, which satisfy 
\begin{eqnarray}
& \varepsilon _{n} u_{n} (\mathbf{r})=[H_{\text{e}}(\mathbf{r}) +U(\mathbf{r})]u_{n} (\mathbf{r})+\Delta (\mathbf{r})v_{n} (\mathbf{r}),\label{eq-3_1}\\
& \varepsilon _{n} v_{n} (\mathbf{r})=-[H_{\text{e}}^{*}(\mathbf{r}) +U(\mathbf{r})]v_{n} (\mathbf{r})+\Delta ^{*} (\mathbf{r})u_{n} (\mathbf{r}).
\label{eq-3_2}
\end{eqnarray}

The two-electron space-spin density matrix (TESSDM)---up to a normalization factor---in the second-quantized representation is \cite{ref4,Lowdin} 
\begin{eqnarray}  
&&\varrho_{\sigma_1 \sigma_2,\sigma'_1 \sigma'_2} (\mathbf{r}_{1} ,\mathbf{r}_{2} ;\mathbf{r}'_{1} ,\mathbf{r}'_{2} )=\nonumber\\ &&~\frac{1}{2} \left\langle \Psi_{\sigma'_2}^{\dag} (\mathbf{r}'_{2} )\Psi_{\sigma'_1}^{\dag} (\mathbf{r}'_{1} ) \Psi_{\sigma_1} (\mathbf{r}_{1} ) \Psi_{\sigma_2}(\mathbf{r}_{2} )\right\rangle. 
\label{eq-4_}
\end{eqnarray} 
On the other hand, the definition of the two-particle Green function is \cite{ref4,Lowdin}
\begin{eqnarray}
&&G_{\sigma_1 \sigma_2,\sigma'_1 \sigma'_2}(\mathbf{r}_1,t_1,\mathbf{r}_2,t_2;\mathbf{r}'_1,t'_1,\mathbf{r}'_2,t'_2)=\\
&&~-\left\langle \mathcal{T}[\Psi _{\sigma_1}(\mathbf{r}_1,t_1) \Psi_{\sigma_2}(\mathbf{r}_2,t_2) \Psi_{\sigma'_2}^{\dag} (\mathbf{r}'_{2},t'_{2}) \Psi_{\sigma'_1}^{\dag } (\mathbf{r}'_{1},t'_{1})]\right\rangle, \nonumber
\label{G-func}
\end{eqnarray}
where $\mathcal{T}$ denotes time-ordering, and all field operators have been written in the Heisenberg picture. Thus the TESSDM is related to the two-particle Green function as follows \cite{ref14,ref15,ref16}:
\begin{eqnarray}
&&\varrho_{\sigma_1 \sigma_2,\sigma_{1'}\sigma_{2'}} (\mathbf{r}_{1},\mathbf{r}_{2};\mathbf{r}'_{1} ,\mathbf{r}'_{2} )=-(1/2)\nonumber\\ && ~\times G_{\sigma_1 \sigma_2, \sigma'_1 \sigma'_2}(\mathbf{r}_1,t_{1},\mathbf{r}_2,t_{2};\mathbf{r}'_1,t^+_{1},\mathbf{r}'_2,t^+_{2}), 
\label{eq-5_} 
\end{eqnarray} 
where $t_i^{+}=t_i+\delta$ (with $\delta\to 0^+$). Furthermore, the two-particle Green function for the superconducting state can be written in terms of single-particle Green functions \cite{ref14,ref15,ref16}
\begin{eqnarray}  
&& G_{\sigma_1 \sigma_2,\sigma'_1 \sigma'_2}(1,2;1',2')=G_{\sigma_1 \sigma'_1}(1,1')G_{\sigma_2 \sigma'_2}(2,2')\nonumber\\
&&-G_{\sigma_1 \sigma'_2}(1,2')G_{\sigma_2 \sigma'_1}(2,1')-F_{\sigma_1 \sigma_2}(1,2)F_{\sigma'_1 \sigma'_2}^{\dag} (1',2'),
\label{eq-6_}
\end{eqnarray} 
where the single-particle Green function $G(i,j)$ and anomalous Green function $F(i,j)$ [here, $i\equiv(\mathbf{r}_i,t_i)$] for the case of $s$-wave superconductors are, respectively, given by \cite{ref16}  
\begin{eqnarray}
 G_{\sigma_{1} \sigma'_{1} } (\mathbf{r}_{1}, t_{1},\mathbf{r}'_{1}, t'_{1})& \equiv & -i\left\langle \mathcal{T}[\Psi_{\sigma_{1}} (\mathbf{r}_{1}, t_{1} )\Psi_{\sigma'_{1}}^{\dag} (\mathbf{r}'_{1}, t'_{1} )]\right\rangle \nonumber\\ &=& \delta _{\sigma_{1} \sigma'_{1} } G(\mathbf{r}_{1}, t_{1};\mathbf{r}'_{1},t'_{1} ),\\
 F_{\sigma_1 \sigma_2}^{\dag} (\mathbf{r}_{1}, t_{1},\mathbf{r}_{2}, t_{2}) &\equiv& -i\left\langle \mathcal{T}[\Psi_{\sigma_1}^{\dag} (\mathbf{r}_{1},t_{1} )\Psi_{\sigma_2}^{\dag} (\mathbf{r}_{2},t_{2} )]\right\rangle\nonumber\\ &=& S_{\sigma_{1} \sigma_{2} } F^{\dag} (\mathbf{r}_{1},t_{1},\mathbf{r}_{2},t_{2}),
\end{eqnarray}
with $S_{\sigma_{1} \sigma_{2} }\equiv i\sigma _{y}$ [recall $\sigma_y=\left(\begin{smallmatrix}0 & -i\\ i & 0 \end{smallmatrix}\right)$ is a Pauli matrix]. If we define time-dependent Bogoliubov coefficients $u_{n} (\mathbf{r},t)$ and $v_{n}(\mathbf{r},t)$ as in the following equations [everything here, except $H_{\text{e}}$, has $(\mathbf{r},t)$ dependence, and henceforth we set $\hbar\equiv c\equiv 1$]:
\begin{eqnarray}
& i\partial_t u_{n}=[H_{\text{e}}+U] u_{n} +\Delta v_{n} ,\label{eq-9_1}\\
&i\partial_t v_{n} =-[H_{\text{e}}^{*} + U]v_{n} + \Delta^{*}u_{n}, 
 \label{eq-9_2} 
\end{eqnarray} 
one can rewrite the Green functions $G_{\uparrow \uparrow } (\mathbf{r}_{1},t_{1},\mathbf{r}'_{1},t'_{1} )\equiv G(\mathbf{r}_{1},t_{1},\mathbf{r}'_{1},t'_{1} )$ and $F_{\uparrow \downarrow }^{\dag } (\mathbf{r}_{1},t_{1},\mathbf{r}_{2},t_{2} )\equiv F^{\dag } (\mathbf{r}_{1},t_{1},\mathbf{r}_{2},t_{2})$, at zero temperature (which we shall assume hereafter), as \cite{ref16}
\begin{eqnarray}  
\hskip-3mm G_{\uparrow \uparrow } (\mathbf{r}_{1} ,t_{1},\mathbf{r}_{2} ,t_{2} ) &=&  i\sum_{n} u_{n} (\mathbf{r}_{1} ,t_{1}) u^*_{n}(\mathbf{r}_{2},t_{2} ) \theta (t_{1} -t_{2} )\nonumber\\
\nonumber\\
&&~ - v^*_{n}(\mathbf{r}_{1},t_{1}) v_{n}(\mathbf{r}_{2} ,t_{2}) \theta(t_{2} -t_{1} ), \\ 
\hskip-3mm F_{\uparrow \downarrow }^{\dag}(\mathbf{r}_{1},t_{1},\mathbf{r}_{2},t_{2} )  &=&  i\sum_{n} u^*_{n}(\mathbf{r}_{2},t_{2}) v_{n}(\mathbf{r}_{1},t_{1}) \theta(t_{1} -t_{2} ) \nonumber\\
&&~+ v_{n}(\mathbf{r}_{2},t_{2}) u^*_{n}(\mathbf{r}_{1},t_{1}) \theta(t_{2} -t_{1}), 
\label{eq-10_}
\end{eqnarray} 
where $\theta(t)$ is the unit step function.

By assuming that the Hamiltonian and the superconducting gap are time-independent, the Bogoliubov coefficients can be written as $u_{n} (\mathbf{r},t)=u_{n} (\mathbf{r})e^{-i\varepsilon _{n} t}$ and $v_{n} (\mathbf{r},t)=v_{n} (\mathbf{r})e^{-i\varepsilon _{n} t}$. We can consider a \textit{weak} magnetic field as a perturbation, which modifies the Bogoliubov coefficients and the energy gap (up to first order) as 
\begin{eqnarray}
X(\mathbf{r})=X^{(0)} (\mathbf{r})+X^{(1)} (\mathbf{r}) +O(X^{(2)}),
\label{pert}
\end{eqnarray}
in which $X\in\{\Delta,u_n , v_n\}$, and superscript ``$(0)$" denotes the solutions to the unperturbed Bogoliubov equations [Eqs.~(\ref{eq-3_1}) and (\ref{eq-3_2})] (i.e., for $\mathbf{A}=0$). Using the London gauge and considering $\mathbf{A}$ not having any component perpendicular to the surface of the superconducting system, one can ignore $\Delta^{(1)}$ \cite{ref16}. We can expand $u_{n}^{(1)} (\mathbf{r})$ in terms of a complete set of normal-state eigenfunctions, and from Eqs.~(\ref{eq-3_1}), (\ref{eq-3_2}), and (\ref{pert}) [or Eqs.~(\ref{eq-9_1}) and (\ref{eq-9_2})], in which case the linearized Bogoliubov equations read
\begin{eqnarray}  
&& \left[\varepsilon _{n} +\frac{\nabla ^{2} }{2m} +\mu -U(\mathbf{r})\right]u_{n}^{(1)} (\mathbf{r})-\Delta^{(0)} (\mathbf{r}) v_{n}^{(1)} (\mathbf{r})  = \nonumber\\
&& \frac{ie}{2m} \left[\mathbf{A}(\mathbf{r})\cdot \nabla +\nabla \cdot \mathbf{A}(\mathbf{r})\right]u_{n}^{(0)} (\mathbf{r}) \\  
&& \left[\varepsilon _{n} -\frac{ \nabla ^{2} }{2m} -\mu +U(\mathbf{r})\right]v_{n}^{(1)} (\mathbf{r})-\Delta^{(0)*} (\mathbf{r})u_{n}^{(1)}(\mathbf{r}) = \nonumber\\
&& \frac{ie}{2m} \left[\mathbf{A}(\mathbf{r})\cdot \nabla +\nabla \cdot \mathbf{A}(\mathbf{r})\right]v_{n}^{(0)} (\mathbf{r}).
\label{eq-13_}
\end{eqnarray} 
We now insert $u_{n}^{(1)} (\mathbf{r})=\sum_{m} a_{nm} \phi _{m} (\mathbf{r}) $, $v_{n}^{(1)} (\mathbf{r})=\sum_{m} b_{nm} \phi_{m} (\mathbf{r})$, $u_{n}^{(0)} (\mathbf{r})=u(\xi _{n} )\phi _{n} (\mathbf{r})$ and $v_{n}^{(0)} (\mathbf{r})=v(\xi _{n} )\phi _{n} (\mathbf{r})$, in which $\phi _{n} (\mathbf{r})$ denotes normal-state eigenfunctions (which can always be taken to be plane waves $\phi _{n} (\mathbf{r})=e^{i\mathbf{k}_{n} \cdot \mathbf{r}}/\sqrt{\Omega}$), $u(\xi _{n}) = \sqrt{(1+\xi _{n}/\varepsilon _{n})/2 }$, $v(\xi_n) = \sqrt{(1-\xi _{n}/\varepsilon _{n})/2}$, $\xi _{n}\equiv n^{2}/(2m)-\mu$, $\varepsilon_{n} \equiv \sqrt{\xi^2_{n} +\Delta^2_{n}}$ (the excitation energy of a quasi-particle with wave vector $\mathbf{n}$), and $\Delta_n=-\sum_{l}\widetilde{V}_{nm}u(\xi_m)v(\xi_m)$ [with the Fourier transform $\widetilde{V}_{nm}\sim \int V(\mathbf{r})e^{-i(\mathbf{n}-\mathbf{m}).\mathbf{r}}~\mathrm{d}^3\mathbf{r}$] \cite{ref16}. Hence, 
\begin{eqnarray}
a_{nm} =\frac{M_{nm} }{\xi _{n}^{2} -\xi _{m}^{2} } \left[(\varepsilon _{n} +\xi _{m} )u(\xi _{n} )+\Delta ^{(0)} v(\xi _{n} )\right],
\label{eq-14_}
\end{eqnarray}
(and similarly for $b_{nm}$ by replacing $u\rightarrow v$ and $\xi_m\rightarrow -\xi_m$) where
\begin{eqnarray}  
M_{nm} & = & 
\frac{-e}{2m} \int \left[\phi _{m}^{*}\phi _{n} \mathbf{k}_{n} +\phi _{n} \phi _{m}^{*} \mathbf{k}_{m} \right]\cdot \mathbf{A} ~\mathbf{d}^3\mathbf{r}.
\label{eq-15_}
\end{eqnarray}
Therefore by calculating the coefficients $a_{nm}$ and $b_{nm}$, we will have $u_{n}^{(1)} (\mathbf{r})$ and $v_{n}^{(1)} (\mathbf{r})$. From Eq.~(\ref{eq-10_}), the linearized single-particle Green function can be rewritten in terms of $u_{n}(\mathbf{r})$ and $v_{n}(\mathbf{r})$, up to first order of perturbation and when $t_{2} \to t_{1}^{+}$,  as follows:
\begin{eqnarray}  
G(\mathbf{r}_{1},\mathbf{r}_{2} ) &\equiv& G(\mathbf{r}_{1},t_{1},\mathbf{r}_{2},t_{2} )\big|_{t_{2} \to t_{1}^{+} } = i\sum _{n}v^*_{n}(\mathbf{r}_{1} )v_{n} (\mathbf{r}_{2} )\nonumber\\&\approx& G^{(0)} (\mathbf{r}_{1},\mathbf{r}_{2} ) + G^{(1)} (\mathbf{r}_{1},\mathbf{r}_{2} ),
\label{eq-16_}
\end{eqnarray}
where
 \begin{eqnarray}
&& G^{(0)}(\mathbf{r}_{1},\mathbf{r}_{2} ) =  i\sum _{n} v_{n}^{(0)*} (\mathbf{r}_{1} )v_{n}^{(0)} (\mathbf{r}_{2}) ,\\
&& G^{(1)}(\mathbf{r}_1,\mathbf{r}_2)= \sum_n v_{n}^{(1)*} (\mathbf{r}_{1} ) v_{n}^{(0)} (\mathbf{r}_{2} ) + v_{n}^{(0)*} (\mathbf{r}_{1} ) v_{n}^{(1)} (\mathbf{r}_{2}).\nonumber\\
\label{eq-16_}
\end{eqnarray} 
More specifically, $G^{(0)}$ (the Green function in the absence of magnetic field \cite{ref4}) and $G^{(1)}$ (the first-order correction) are given as
\begin{eqnarray} 
&&\hskip-3mm G^{(0)} (\mathbf{r}_{1} ,\mathbf{r}_{2} )\equiv G^{(0)} (r)\nonumber
\\&&  =\frac{-im}{2\pi ^{2} r} \int _{0}^{\infty }v^2(\xi_{k}) \sin (kr)k~\mathrm{d}k
\\&& = G^{(0)}(0) \frac{\sin(k_F r)}{2k_F r} \Bigl[2-\pi I_1\Bigl( \frac{r}{\pi\xi_0}\Bigr)+\pi L_1\Bigl( \frac{r}{\pi\xi_0}\Bigr) \Bigr],  
\label{eq-18_} \\
&&\hskip-3mm G^{(1)}(\mathbf{r}_{1},\mathbf{r}_{2} ) = i\sum_{nm}  b_{nm} v^{*} (\xi _{n} )\phi^*_{n}(\mathbf{r}_{1} )\phi _{m} (\mathbf{r}_{2} ) \nonumber\\
&&\hskip21mm~ + b^*_{nm}v(\xi _{n} )\phi _{n} (\mathbf{r}_{2} )  \phi^*_{m}(\mathbf{r}_{1} ),
\label{eq-17_} 
\end{eqnarray} 
in which $\mathbf{r}\equiv \mathbf{r}_{1} -\mathbf{r}_{2}$ is the relative distance of electrons in a Cooper pair, $I_1(x)$ is the order $1$ modified Bessel function, and $L_1(x)$ is the order $1$ modified Struve function \cite{Abram}. At the continuum limit ($\mathbf{k}_{n} \to \mathbf{k}$, $\mathbf{k}_{m} \to \mathbf{k}'$, $(1/\Omega) \sum_n \to [1/(2\pi)^{3}]  \int\mathrm{d}^3\mathbf{k} $), one can obtain
\begin{eqnarray}  
G^{(1)} (\mathbf{r}_{1},\mathbf{r}_{2} )& =& -\frac{i\Omega}{2(2\pi )^{6} } \int Y_{kk'} M_{kk'} \label{eq-24_-}\\
&&~\times \bigl[e^{i(\mathbf{k}'\cdot \mathbf{r}_{2} - \mathbf{k}\cdot \mathbf{r}_{1})} +\chi  e^{i(\mathbf{k}\cdot \mathbf{r}_{2} - \mathbf{k}'\cdot \mathbf{r}_{1})} \bigr] ~\mathrm{d}^3 \mathbf{k}\mathrm{d}^3\mathbf{k}', \nonumber
\end{eqnarray} 
where $Y_{kk'} = \bigl[(\varepsilon  -\xi  )(\varepsilon  -\xi' )+\Delta^{2}\bigr]/[\varepsilon (\xi^2-{\xi'}^2)]$, $M_{kk'}$ is the continuum version of Eq.~(\ref{eq-15_}), and $\chi$ is $+1$ ($-1$) if $M_{kk'}$ is real (imaginary).
\begin{figure}[tp]
\includegraphics[scale=.7]{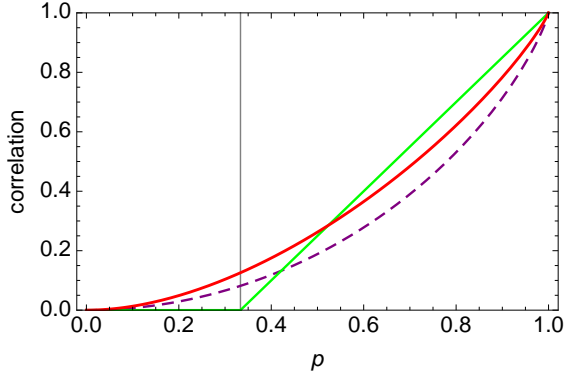}
\caption{(Color online) Quantum and classical correlations (red, solid line; purple, dashed line, respectively) as functions of $p$ of the Cooper pair density matrix. Here, entanglement is calculated through concurrence (green, solid line), and quantum correlation is calculated through quantum discord (which already excludes classical correlations from quantum mutual information)---Eq.~(\ref{Q-dis}).}
\label{fig:fig-r}
\end{figure}

Additionally, we have 
\begin{eqnarray} 
F^{\dag } (\mathbf{r}_{1},\mathbf{r}_{2}) &\equiv& F^{\dag } (\mathbf{r}_{1} ,t_{1} ,\mathbf{r}_{2} ,t_{2} )\big|_{t_{2} \to t^+_{1}}  =i\sum_{n}v_{n} (\mathbf{r}_{2})u^*_{n}(\mathbf{r}_{1} )\nonumber\\
& \approx& F^{(0)\dag} (\mathbf{r}_{1},\mathbf{r}_{2} ) + F^{(1)\dag} (\mathbf{r}_{1},\mathbf{r}_{2} ), 
\label{eq-19_} 
\end{eqnarray}
where
\begin{eqnarray}
F^{(0)\dag} (\mathbf{r}_{1},\mathbf{r}_{2} )&=&i\sum_{n} v_{n}^{(0)} (\mathbf{r}_{2} )u_{n}^{(0)*} (\mathbf{r}_{1} ) \label{eq-19_1}\\ 
F^{(1)\dag} (\mathbf{r}_{1},\mathbf{r}_{2} )&=&i\sum_{n}v_{n}^{(1)} (\mathbf{r}_{2}) u_{n}^{(0)*} (\mathbf{r}_{1}) + v_{n}^{(0)} (\mathbf{r}_{2} )u_{n}^{(1)*} (\mathbf{r}_{1} ), \nonumber\\
\label{eq-19_} 
\end{eqnarray} 
or equivalently
\begin{eqnarray}  
&&\hskip-3mm F^{\dag (0)} (\mathbf{r}_{1},\mathbf{r}_{2} ) =  i\sum_{n} v(\xi _{n} )\phi _{n} (\mathbf{r}_{2} ) u^{*} (\xi _{n} )\phi^*_{n}(\mathbf{r}_{1})\nonumber\\
&& ~\equiv F^{(0)*} (r) =\frac{i}{\pi^2 r}\int\frac{\sin(kr)k}{\sqrt{1+(\xi_{k}/\Delta_{k})^2}}~\mathrm{d}k\nonumber\\
&&~\approx i\Delta N(0) \frac{\sin(k_{F} r)}{k_{F} r}K_{0}\Bigl(\frac{r}{\pi \xi _{0}}\Bigr),\label{eq-20_1}\\
&&\hskip-3mm F^{(1)\dag } (\mathbf{r}_{1},\mathbf{r}_{2} )  =  i\sum_{nm} b_{nm} u^{*} (\xi _{n} )\phi^*_{n}(\mathbf{r}_{1} )\phi_{m}(\mathbf{r}_{2} )\nonumber\\
&&\hskip21mm~ + a^*_{nm} v(\xi _{n} ) \phi _{n}(\mathbf{r}_{2} ) \phi^*_{m}(\mathbf{r}_{1} ), 
\label{eq-20_}
\end{eqnarray}
where $N(0)\equiv (1/2)\bigl[k^{2} (\mathrm{d}k/\mathrm{d}\varepsilon _{k})\bigr]_{\varepsilon _{k} =\varepsilon _{F} } =m k_{F}/(2\pi^2)$ is the density of states for one spin projection at the Fermi surface \cite{ref4}, and $K_{0} (y)$ is the order $0$ modified Bessel function \cite{Abram}, and in Eq.~(\ref{eq-24_-}) and here $\Delta$ is $\Delta^{(0)}$ (constant for $s$-wave superconductors). At the continuum limit, one can also obtain
\begin{eqnarray} 
&&F^{(1)\dag} (\mathbf{r}_{1} ,\mathbf{r}_{2} )  =  \frac{i \Omega\Delta }{2(2\pi )^{6} } \int  \mathrm{d}^3\mathbf{k}\mathrm{d}^3\mathbf{k}' \frac{M_{kk'}}{\xi ^{2} -\xi '^{2}}\Big\{e^{-i(\mathbf{k}\cdot\mathbf{r}_{1} - \mathbf{k}'\cdot \mathbf{r}_{2})}\nonumber\\ &&\Bigl[1+\frac{(\varepsilon-\xi')\Delta}{\varepsilon^2}+\frac{\xi}{\varepsilon}\Bigr]+\chi e^{i(\mathbf{k}\cdot\mathbf{r}_{2} - \mathbf{k}'\cdot\mathbf{r}_{1})} \Bigl[1+\frac{(\varepsilon-\xi')\Delta}{\varepsilon^2}-\frac{\xi}{\varepsilon}\Bigr]\Big\}.\nonumber\\
\label{eq-36_} 
\end{eqnarray}

The TESSD, for the case $(\mathbf{r}_{1},\mathbf{r}_{2}) = (\mathbf{r}'_{1},\mathbf{r}'_{2})$ and up to first order approximation, reads \cite{ref4}
\begin{widetext}
\begin{eqnarray}  
\varrho_{\sigma_{1} \sigma_{2},\sigma'_{1} \sigma'_{2} }(\mathbf{r}_{1},\mathbf{r}_{2} ) & = & -(1/2)\Bigl[\delta _{\sigma_{1} \sigma'_{1} } \delta _{\sigma'_{2} \sigma'_{2} } \{G^{(0)2} (0)+G^{(0)} (0 )G^{(1)} (\mathbf{r}_{2} ,\mathbf{r}_{2} ) + G^{(0)}(0)G^{(1)} (\mathbf{r}_{1} ,\mathbf{r}_{1}) \} \nonumber \\ && -\delta _{\sigma_{1} \sigma'_{2} } \delta _{\sigma_{2} \sigma'_{1} } \{G^{(0)2} (\mathbf{r})+G^{(0)} (\mathbf{r})G^{(1)} (\mathbf{r}_{2} ,\mathbf{r}_{1} )+ G^{(0)} (\mathbf{r} ) G^{(1)} (\mathbf{r}_{1} ,\mathbf{r}_{2} )\} \nonumber \\  &&-S_{\sigma_{1} \sigma_{2} } S_{\sigma'_{1} \sigma'_{2} } \{|F^{(0)} (r)|^{2} +F^{(0)}(r) F^{(1)*} (\mathbf{r}_{1} ,\mathbf{r}_{2} )+ F^{(0)*} (r) F^{(1)} (\mathbf{r}_{1} ,\mathbf{r}_{2} )\}\Bigr].
\label{eq-41_}
\end{eqnarray} 
\end{widetext}
Using $n(\mathbf{r})=-i\mathrm{Tr}[G(\mathbf{r},t,\mathbf{r},t^{+} )]$, we can write $G^{(1)} (\mathbf{r}_{1} ,\mathbf{r}_{1} )=-in(\mathbf{r}_{1} )/2 $ (similarly for $n(\mathbf{r}_2)$) and $G^{(0)} (0)\equiv -in/2$, where $n$ is the electron density. We thus obtain 
\begin{eqnarray}  
\widetilde{\varrho}\equiv \frac{\varrho}{\mathrm{Tr}[\varrho]} =\frac{1}{2(\rho_1+\rho_2)} \left(\begin{array}{cccc}  \rho_1 &  &  &  \\  &  \rho_2 & \rho_3 &  \\  & \rho_3 & \rho_2 &  \\  &  &  & \rho_1 \end{array}\right), 
\label{eq-42_}
\end{eqnarray} 
where
\begin{eqnarray}
\rho_1 & = & d+q+ (1-g^{2} )n^2/8,\\
\rho_2 & = & d+w+ (1+f^{2} )n^2/8,\\
\rho_3 & = & q-w-(g^{2} +f^{2} )n^2/8,
\end{eqnarray}
with $f(r)\equiv F^{(0)}(r)/G^{(0)} (0)$, $g(r)\equiv G^{(0)}(r)/G^{(0)}(0)= 2iG^{(0)} (r)/n$ \cite{ref4}, $d(\mathbf{r}_1,\mathbf{r}_2)=[n(\mathbf{r}_{1} )+n(\mathbf{r}_{2} )]n/8$, $q(\mathbf{r}_1,\mathbf{r}_2)=-i[G^{(1)}(\mathbf{r}_{2},\mathbf{r}_{1} )+G^{(1)} (\mathbf{r}_{1},\mathbf{r}_{2} )]g(r)n/4$, and $w(\mathbf{r}_1,\mathbf{r}_2)=\mathrm{Re}[F^{(0)} (r)F^{(1)* }(\mathbf{r}_{1},\mathbf{r}_{2} )]$ \cite{Herm}. It is straightforward to see that $\widetilde{\varrho}$ is in fact a Werner state \cite{ref4,ref11} 
\begin{eqnarray}  
\widetilde{\varrho} =\frac{1-p}{4}\openone+p | \psi^{-} \rangle \langle \psi^{-}|,  
\label{eq-44_}
\end{eqnarray} 
where $\openone$ is the $4\times 4$ identity matrix, $| \psi ^{-} \rangle =(| \uparrow\downarrow\rangle -| \downarrow\uparrow\rangle)/\sqrt{2}$ is the spin singlet, and 
\begin{eqnarray}
p=\frac{w-q+(f^2 + g^2)n^2/8}{2d+w+q+(f^2 -g^2+2)n^2 /8}. 
 \label{eq-45_} 
\end{eqnarray} 

It is interesting to see how the state (and the correlations thereof) vary due to the application of the magnetic field (cf. Ref.~\cite{ref4}). Neglecting $f(r)$ \cite{ref4} and terms higher than $O\bigl(G^{(1)}\bigr)$, the perturbed $p$ [denoted below as $p^{(1)}$] has the following form in terms of the unperturbed one which is for the case without magnetic field [denoted below as $p^{(0)}$]:  
\begin{eqnarray}
\hskip-3mm p^{(1)} \approx p^{(0)} -2\frac{g^2(r)d(\mathbf{r}_1,\mathbf{r}_2)+q(\mathbf{r}_1,\mathbf{r}_2)}{[2-g^2(r)]^2}. 
\label{p0-p1}
\end{eqnarray}

As a result, it can be argued that for some values of the magnetic field, $p^{(1)}\approx p^{(0)}$. This implies that the addition of a nonvanishing, weak magnetic field sometimes does not alter the related density matrix (and hence, as we see below, its quantum correlations), while in general it affects the density matrix as in Eq.~(\ref{p0-p1}). 


Having obtained the form of the density matrix, it is usually of interest to know how it is correlated quantum mechanically. To that aim, entanglement is usually considered as a measure of quantum correlation. Using the properties of a Werner state \cite{Xstates}, it is straightforward to calculate entanglement through, for example, the Peres-Horodecki separability criterion \cite{ref12}. Specifically, a $2\times 2$ Werner state is entangled if and only if $p>1/3$ \cite{ref3,ref4,ref12}. 

Although non-separability (i.e., entanglement) is considered to be a purely quantum mechanical effect, it has been known that it is not responsible for all quantum correlations. To quantify non-classicality of correlations, ``quantum discord" has been introduced as a finer measure. It is defined to be the difference between two classically identical expressions for the mutual information \cite{Zurek}. While no separable state can have quantum entanglement, some of them may exhibit nonvanishing quantum discord. For Werner states of the form (\ref{eq-44_}), it has been shown that any $p>0$ implies a nonvanishing discord \cite{Zurek,Xstates}. More explicitly, quantum discord can be obtained with the following relation \cite{Xstates}:
\begin{eqnarray}
Q(\widetilde{\varrho}) & = & \frac{1}{4}\Bigl[(1-p)\log_2(1-p) +(1-3p)\log_2(1+3p)\nonumber\\
&&~ -2(1+p)\log_2(1+p) \Bigr],
\label{Q-dis}
\end{eqnarray}
which can be expanded based on Eq.~(\ref{p0-p1}) to see the effect of the magnetic field. Figure~\ref{fig:fig-r} depicts quantum discord and concurrence \cite{Wootters} (as a specific measure of entanglement, equal to $\max\{0,(3p-1)/2\}$) for the spin state of a Cooper pair. 


In summary, we considered an $s$-wave superconductor subjected to a weak magnetic field, and obtained spin-space density matrix of electrons of a Cooper pair. We employed the relation of the density matrix with the Green functions, and  applied perturbation theory to solve the corresponding Bogoliubov equations. This density matrix was found to be in the form of a Werner state, hence it is not necessarily maximally entangled. This can be attributed to the fact that correlations between electrons in the ground state of a superconducting state are not localized in each Cooper pair \cite{ref4}. Nonetheless, a nonvanishing quantum correlation may have some relevant implications on the overall quantum correlation in the system. We thus calculated quantum discord---as a measure of quantum correlation---for the obtained density matrix, and showed how a weak magnetic field affects this correlation. Besides basic importance of such calculations, we hope that our study can shed some light on potential applications of superconductors in, for example, quantum information science through nanoscale systems \cite{Nori}.



\begin{thebibliography}{99}

\bibitem{ref1} 
 T. J. Osborne and M. A. Nielsen, Phys. Rev. A \textbf{66}, 032110 (2002); A. Osterloh, L. Amico, G. Falci, and R.  Fazio, Nature  \textbf{416}, 608 (2002); L. Amico, R. Fazio, A. Osterloh, and V. Vedral, Rev. Mod. Phys. \textbf{80}, 517 (2008).


\bibitem{ref2} F. Verstraete, M. Popp, and J. I. Cirac, Phys. Rev. Lett. \textbf{92}, 027901 (2004).

\bibitem{Nielsen:book} M. A. Nielsen and I. L. Chuang, \textit{Quantum Information and Quantum Computation} (Cambridge University Press, Cambridge, U.K., 2000).

\bibitem{ref3} S. Oh and J. Kim, Phys. Rev. A \textbf{69}, 054305 (2004).

\bibitem{ref4} S. Oh, and J. Kim, Phys. Rev. B \textbf{71}, 144523 (2005).

\bibitem{Hubbard} S.-J. Gu, S.-S. Deng, Y.-Q. Li, and H.-Q. Lin, Phys. Rev. Lett. \textbf{98}, 086402 (2004).

\bibitem{Zurek} H. Ollivier and W. H. Zurek, Phys. Rev. Lett. \textbf{88}, 017901 (2001).

\bibitem{ref14} A. L. Fetter and J. D. Walecka, \textit{Quantum Theory of Many-Particle Systems} (McGraw-Hill, New York, 1971). 

\bibitem{ref15} A. A. Abrikosov, L. P. Gorkov, and P. E. Dzyaloshinski, \textit{Methods of Quantum Field Theory in Statistical Physics} (Dover Publications, New York, 1975).

\bibitem{ref16} J. B. Ketterson and S. N. Song, \textit{Superconductivity} (Cambridge University Press, U.K., 1995).

\bibitem{Lowdin} P.-O. L\"{o}wdin, Phys. Rev. \textbf{97}, 1490 (1955).

\bibitem{Abram} M. Abramowitz and I. A. Stegun, \textit{Handbook of Mathematical Functions with Formulas, Graphs, and
Mathematical Tables} (Dover Publications, New York, 1972).










\bibitem{Herm} The Hermiticity of $\varrho$ [$\varrho_{\sigma_1\sigma_2,\sigma'_1\sigma'_2}(\mathbf{r}_1,\mathbf{r}_2)=\varrho^*_{\sigma'_1\sigma'_2,\sigma_1\sigma_2}(\mathbf{r}_1,\mathbf{r}_2)$] is guaranteed for the Hermiticity of $G$ [Eq.~(\ref{eq-6_})], which translates into $G^{(1)}(\mathbf{r}_1,\mathbf{r}_2)=-G^{(1)*}(\mathbf{r}_2,\mathbf{r}_1)$ and similarly for $F^{(1)\dag}(\mathbf{r}_1,\mathbf{r}_2)$.


\bibitem{ref11} R. F. Werner, Phys. Rev. A \textbf{40}, 4277 (1989).

\bibitem{ref12} A. Peres, Phys. Rev. Lett. \textbf{77}, 1413 (1996); M. Horodecki, P. Horodecki, and R. Horodecki, Phys. Lett. A \textbf{223}, 1 (1996).

\bibitem{Xstates} M. Ali, A. R. P. Rau, and G. Alber, Phys. Rev. A \textbf{81}, 042105 (2010).

 \bibitem{Wootters} W. K. Wootters, Phys. Rev. Lett. \textbf{80}, 2245 (1998).
 
 \bibitem{Nori} J. Q. You and F. Nori, Phys. Tod. \textbf{58}, 42 (2005).

\end{thebibliography}
\end{document}